# Net-Zero Settlement in Distribution Markets

Sina Parhizi, Alireza Majzoobi, and Amin Khodaei
Department of Electrical and Computer Engineering
University of Denver
Denver, Colorado
sina.parhizi@du.edu, alireza.majzoobi@du.edu, amin.khodaei@du.edu

*Abstract—* **Introduction of market mechanisms in distribution systems is currently subject to extensive studies. One of the challenges facing Distribution Market Operators (DMOs) is to implement a fair and economically efficient pricing mechanism that can incentivize consumers to positively contribute to grid operations and to improve economic performance of the distribution system. This paper studies a penalty-based pricing mechanism in distribution markets and further investigates the interrelationship between the locational marginal prices (LMPs) at transmission and distribution levels. As a result, a closed-form relationship between these LMPs is derived. The possibility of zeroing out the settlement profit is further investigated under the proposed pricing mechanism.**

## NOMENCLATURE

*Indices:*

| | |
|---|---|
| $t$ | Index for hours. |
| $m$ | Index for proactive customers. |
| $l$ | Index for lines. |

*Parameters:*

| | |
|---|---|
| $\mu$ | Penalty factor for the power deviation. |
| $\lambda^T$ | Transmission locational marginal price (T-LMP). |
| $\lambda^D$ | Distribution locational marginal price (D-LMP). |
| $H$ | Generation shift factor. |

*Variables:*

| | |
|---|---|
| $C^c$ | Customer payments to the DMO. |
| $C^u$ | DMO payment to the ISO. |
| $C^\Delta$ | DMO cost surplus. |
| $\Delta P^M$ | Power deviation. |
| $D$ | Adjustable load. |
| $D^f$ | Fixed load. |
| $F(P,I)$ | Dispatchable unit operation cost. |
| $\pi$ | Dual variable for the power balance equation. |
| $P$ | DER output power. |
| $P^M$ | Power transfer to the microgrid. |
| $PD^M$ | Total assigned power from the main grid. |
| $PL$ | Line power flow. |
| $\tau^-, \tau^+$ | Dual variables for line flow limits. |
| $\alpha, \beta$ | Dual variables for demand limits. |

## I. INTRODUCTION

DISTRIBUTION SYSTEMS are currently subject to widespread attention as part of the agendas outlining the utilities of the future [1], seeking to reform, modernize, and upgrade the electric power system infrastructure. Factors such as economic requirements for higher efficiency, compliance with environmental regulations, and need for higher reliability and resiliency are among major drivers of these plans [2]. Utilization of modern communications and control technologies, providing additional services to customers, and improving grid monitoring and protection are examples of the goals to be investigated and promoted in these initiatives. The efficient operation of the distribution system is also among the objectives sought in the agendas proposed for the future of electric power systems [1]; a future that is deemed to witness significant growth in the participation of microgrids, electric vehicles and distributed energy resources (DER) in the distribution level [3], [4], a new path forward for distribution systems who have traditionally been radial and have had unidirectional power flowing in their lines. An optimally operated distribution system would create incentives for the customers to consume electricity taking into account economic considerations. Thus, growing trend of responsive and proactive consumers in distribution systems can be considered as another driving force for reforming the structure and operation of these systems [5].

Market-based operation of distribution systems is one of the initiatives put forth in the power industry aiming for fostering competition through creating market mechanisms in distribution system [6]–[9]. Optimal scheduling of microgrids [10]–[12], demand response [13] and DERs in a market environment has been under extensive studies. One of the challenges facing the distribution market operation, which is yet to be effectively addressed, is pricing of electricity in those systems. A pricing mechanism reflecting the operating condition of the grid has been discussed in [14] to improve the economic efficiency of the distribution system operation. Such a mechanism would reflect the actual state of the distribution system as well as utility grid economic realities. At the distribution system, it is necessary for the selected pricing mechanism to adapt to features such as radial network topology, presence of feeders and proactive consumers, as well as the connection to the utility grid, so as to ensure a fair and economical settlement of the system operation costs. The similarities and differences between the wholesale and distribution pricing mechanisms are highlighted in [15]. It is demonstrated that transmission LMPs

(T-LMP), line congestions because of high loads, and distribution system losses, can all influence distribution LMPs (D-LMP). Locational marginal pricing has been used by the ISOs to determine the electricity prices at any certain bus within the high voltage power system [16]. Increasing penetration of proactive customers, including responsive consumers, prosumers, and microgrids, in the distribution system makes it necessary for a realistic measure reflecting the cost of energy delivery at different locations in the distribution system [17]. Such a pricing mechanism would provide a viable means to assess the economic performance of different customers in the system.

There have been several research works related to locational marginal pricing in the price-based scheduled distribution systems. Reference [18] has introduced nodal pricing in such a distribution system considering the marginal loss, while ignoring network congestion. Contribution of DERs to the loss reduction in the distribution system is used in [19] to calculate the D-LMPs by an iterative approach based on neural networks. Demand responsive loads have been considered in [20] when calculating D-LMPs using an AC optimal power flow in maximizing the social welfare. Distribution pricing has been carried out in a hierarchical dispatch model in [21], which offers a decentralized approach to perform the dispatch by various regional operators. A variety of methods based on quadratic programming for calculating loss factors to be used in D-LMP calculation is offered in [22]. A distribution system market with high penetrations of solar generation is proposed in [23], where D-LMPs are calculated as a function of the upstream T-LMP and losses. In [24], an iterative model has been used to solve the optimal power flow to obtain the D-LMPs, simultaneously considering distribution and transmission systems. A distributed scheme for calculating D-LMPs based on T-LMPs with participation factors is presented in [25], where D-LMPs are used to drive distribution system controls. The calculation of the D-LMPs in highly unbalanced distribution systems is a challenge addressed in [26], in which a method is presented to calculate loss sensitivities in the distribution system. An iterative method is presented in [27] to calculate stochastic D-LMPs based on point estimate method considering greenhouse gas emissions. Optimal power flow based on three-phase Current Injection Method is used in [28] to calculate D-LMPs and address the concerns raised by using AC and DC OPF. A game-theoretic approach is proposed in [29] to clear the distribution electricity market and calculate D-LMPs. Another important issue to be considered in distribution markets is the market settlement, i.e., how to calculate the payments of consumers as well as payments to suppliers. In the transmission system, the settlement will be net-zero if there is no congestion, i.e., the total payment by consumers will be paid to suppliers (generation companies). If there is congestion, however, additional charges will be imposed to consumers that cause congestion (calculated inherently in T-LMP as a congestion cost term), thus resulting in a nonzero settlement. ISOs, however, are equipped with additional market mechanisms, such as Financial Transmission Rights (FTR) and Auction Revenue Right (ARR) to allocate the obtained revenue among market participants [30]. Such mechanisms do not exist in the distribution system, hence highlighting the need to find a mechanism to efficiently redistribute this revenue.

Presence of an entity that centrally administers transactions in the distribution system has been considered essential for the distribution systems to be able to set up a locational pricing mechanism similar to that in transmission system [31]. The Distribution Market Operator (DMO) is an entity described in [11] and [16] that has the capability to clear the distribution market and determine D-LMPs. The clearing and settlement procedures for a distribution market operated by the DMO are investigated in [32]. Similar concepts have been put forth in the literature. The Distribution System Operator (DSO), as described in [7], is an entity that in its maximalist extent can facilitate a retail market for demand response and DERs. State of New York is pursuing a concept called Distribution Service Platform Provider (DSPP) that would be responsible for enabling active customer participation in the distribution market [6]. Furthermore, the Independent Distribution System Operator (IDSO) as defined in [9] would be responsible for promoting market mechanisms to integrate DERs.

This paper first develops an analytical relationship between the D-LMPs and the upstream T-LMP with a focus on high penetration of proactive customers. These D-LMPs, which can be adjusted using a defined penalty factor, are used to zero out any settlement revenue made by the DMO and thus accordingly address the challenge of mismatch between the revenue made from the customers and the payment to the upstream network.

The rest of the paper is organized as follows. The proposed formulation for the distribution system pricing model and the net-zero settlement model are presented in Sections II and III, respectively. Case studies are presented in Section IV. The paper is concluded in Section V.

## II. PRICING MODEL

The distribution market in this work is considered to be composed of proactive customers, each equipped with a certain amount of flexible loads and DERs. It is assumed that at each bus in the distribution system there is at least one proactive customer that submits demand bids to the DMO, hence index $m$ is used to denote both buses and customers. The DMO receives the bids and submits one aggregated bid to the ISO, which the ISO would accordingly use to clear the market and determine awards for each market player, including the DMO. Once the DMO is notified of the amount of awarded power, it will assign this power to distribution customers based on their submitted bids and prevailing operational constraints. The operation cost of the distribution market is defined as in (1):

$$\min\left(-\sum_t\sum_m F_m(D_{mt})+\sum_t\lambda_t^T P_t^M+\sum_t\mu\,|\,\Delta P_t^M\,|\right) \quad (1)$$

This objective function minimizes the power purchase from the utility grid while maximizing the benefits of loads in the distribution system, i.e., it seeks to maximize the social welfare at the distribution market.

A penalty term is added to the objective function to minimize the deviation between the power assigned by the ISO

(i.e., the assigned power) and the power that is to be purchased from the utility via clearing the market (i.e., the scheduled power). The power deviation is calculated as $\Delta P_t^M = P_t^M - PD_t^M$. Fig. 1 depicts a sample demand bid showing fixed load, maximum load, and assigned and scheduled powers. The objective is subject to the following operational constraints:

$$\sum_m D_{mt}^f + \sum_m D_{mt} = P_t^M \qquad\qquad \forall t \tag{2}$$

$$-D_{mt} \leq 0 \qquad\qquad \forall t, \forall m \tag{3}$$

$$D_{mt} \leq D_m^{\max} \qquad\qquad \forall t, \forall m \tag{4}$$

$$-H_{l0}P_t^M + \sum_m H_{lm}(D_{mt}^f + D_{mt}) \leq PL_l^{\max} \ \forall t, \forall l \tag{5}$$

$$H_{l0}P_t^M - \sum_m H_{lm}(D_{mt}^f + D_{mt}) \leq PL_l^{\max} \qquad \forall t, \forall l \tag{6}$$

These constraints represent the distribution system load balance (2), customer bids minimum and maximum limits (3)-(4), and line flow limits based on shift factors (5)-(6). The first term in (5)-(6) accounts for the shift factor of distribution system power transfer with the upstream network. The objective function can be written as a function of only net load values by substituting (2) in (1). The proposed model considers a radial structure for the distribution network, thus flow of each line can be represented as a function of nodal injections. It is further assumed that the microgrids are in close proximity, thus active losses are small compared to the power transferred through the distribution lines and therefore are negligible. The formulation of the active/reactive power losses in distribution networks encompassing a larger area will be employed in the future works.

Depending on the values of scheduled and assigned powers, the sign of the deviation term within the absolute function can be determined. If $P_t^M < PD_t^M$, then $|\Delta P_t^M|$ would be equal to $-\Delta P_t^M$. On the other hand, if $P_t^M > PD_t^M$, then $|\Delta P_t^M|$ would be equal to $+\Delta P_t^M$. The dual variables of constraints (2)-(6) are denoted respectively by $\pi$, $\alpha$, $\beta$, $\tau^-$, and $\tau^+$. After updating the objective with substituting the $|\Delta P_t^M|$, the Lagrangian function for this optimization problem can be written as follows (both signs for $|\Delta P_t^M|$ are considered):

$$
\begin{aligned}
L = &-\sum_t \sum_m F_m(D_{mt}) + \sum_t \sum_m \lambda_t^T P_t^M \mp \sum_t \mu \Delta P_t^M \\
&+ \pi_t(\sum_m D_{mt}^f + \sum_m D_{mt} - P_t^M) \\
&+ \sum_l \tau_{lt}^-(-H_{l0}P_t^M + \sum_m H_{lm}(D_{mt}^f + D_{mt}) - PL_l^{\max}) \\
&+ \sum_l \tau_{lt}^+(H_{l0}P_t^M - \sum_m H_{lm}(D_{mt}^f + D_{mt}) - PL_l^{\max}) \\
&+ \sum_m \alpha_{mt}(-D_{mt}) \\
&+ \sum_m \beta_{mt}(D_{mt} - D_m^{\max})
\end{aligned}
\tag{7}
$$

Taking the derivatives of the Lagrangian function, the following equations are obtained:

$$\frac{\partial L}{\partial D_{mt}} = \lambda_{mt}^D = -\frac{\partial F_m}{\partial D_{mt}} + \pi_t + \sum_l (\tau_{lt}^- - \tau_l^+)H_{lm} - \alpha_{mt} + \beta_{mt} \tag{8}$$

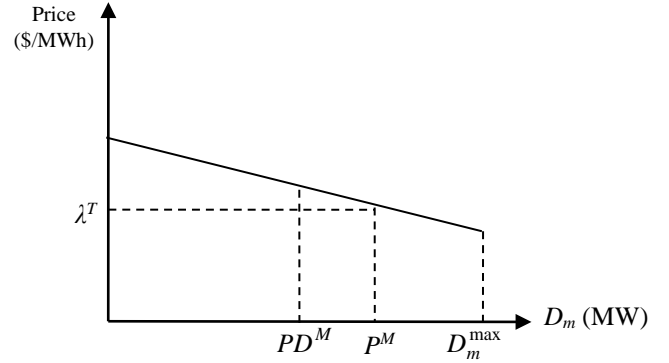

Figure 1. Demand bid curve for customer at bus $m$. $F_m(D_m)$ for any $D_{mt}$ is the area under this curve.

$$\frac{\partial L}{\partial P_t^M} = \lambda_t^T \mp \mu - \pi_t - \sum_l (\tau_{lt}^- - \tau_l^+)H_{l0} = 0 \tag{9}$$

By obtaining $\pi$ from (9) and substituting in (8), the following equation among dual variables is obtained:

$$\lambda_{mt}^D = \lambda_t^T - \frac{\partial F_m}{\partial D_{mt}} \mp \mu + \sum_l (H_m + H_{l0})(\tau_l^+ - \tau_l^-) - \alpha_{mt} + \beta_{mt} \tag{10}$$

The D-MLPs can be calculated from (10). It is observed that the D-LMP is a function of not only upstream T-LMP but also distribution line congestions, the power deviation penalty, and the demand bid submitted by the participants. As pointed out in [15], in case the distribution system is unable to supply its loads using the utility grid power purchases, D-LMPs will solely be determined by the characteristics of distribution system topology and its local consumers.

## III. Market Settlmenet

Using the derived D-LMPs, the DMO can efficiently calculate the cost of power purchase from the ISO as well as the payment of its customers. These two costs are shown in (11) and (12), respectively.

$$C^u = \sum_t \lambda_t^T P_t^M = \sum_t \sum_m \lambda_t^T (D_{mt}^f + D_{mt}) \tag{11}$$

$$C^c = \sum_t \sum_m \lambda_{mt}^D (D_{mt}^f + D_{mt}) \tag{12}$$

In order to have a net-zero settlement, these two costs should be equal (13).

$$C^\Delta = C^c - C^u = \sum_t \sum_m (\lambda_{mt}^D - \lambda_t^T)(D_{mt}^f + D_{mt}) = 0 \tag{13}$$

Using (10), it can be written as:

$$C^\Delta =$$

$$\sum_t \sum_m (\frac{\partial F_m}{\partial D_{mt}} \mp \mu + \sum_l (H_m - H_{l0})(\tau_l^+ - \tau_l^-) - \alpha_{mt} + \beta_{mt})(D_{mt}^f + D_{mt}) \tag{14}$$

By equating this equation to zero, $\mu$ can be set so that settlement mismatch is removed. The sign of $\mu$ in (14) would be determined by the value of the assigned power to the DMO by the ISO.

## IV. NUMERICAL STUDIES

The pricing model is applied to the IEEE 13-bus test system (shown in Fig. 2) [33]. The distributed energy resource data, utilized by proactive customers to offer a responsive load, are borrowed from [32]. It is assumed that customers submit step-wise demand bids into the distribution market, however here a single-step bid is considered. The bid prices reflect the incremental generation costs of local DGs. The maximum daily fixed and adjustable loads for all proactive customers are shown in Table I. Table II shows the transmission LMP, which is the marginal electricity price at bus 1 where this distribution system is connected to the upstream transmission system. Table III presents the amount of awarded power from the ISO to the DMO, which is determined through the wholesale market.

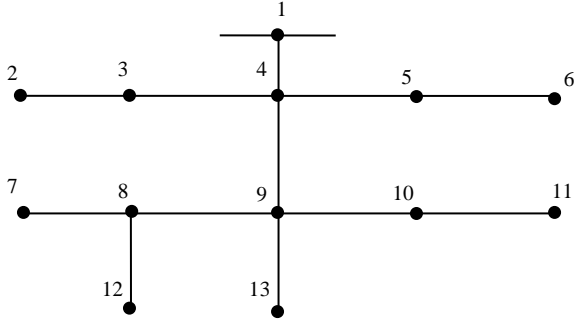

Figure 2. IEEE 13-bus standard test system.

TABLE I
SPECIFICATION OF FIXED AND ADJUSTABLE LOADS IN TEST SYSTEM

| Bus | 2 | 3 | 4 | 5 | 6 | 7 |
|---|---|---|---|---|---|---|
| Peak load (MW) | 0.86 | 0.69 | 0.00 | 0.00 | 1.28 | 0.86 |
| Max. adj. load (MW) | 0.69 | 0.00 | 0.00 | 0.00 | 0.00 | 0.69 |
| Bus | 8 | 9 | 10 | 11 | 12 | 13 |
| Peak load (MW) | 0.00 | 2.14 | 1.28 | 1.71 | 0.86 | 0.00 |
| Max. adj. load (MW) | 0.00 | 1.07 | 0.00 | 1.21 | 0.59 | 0.00 |

TABLE II
HOURLY TRANSMISSION LMP

| Time (h) | 1 | 2 | 3 | 4 | 5 | 6 | 7 | 8 |
|---|---|---|---|---|---|---|---|---|
| Price ($/MWh) | 15.03 | 10.97 | 13.51 | 15.36 | 18.51 | 21.80 | 17.30 | 22.83 |
| Time (h) | 9 | 10 | 11 | 12 | 13 | 14 | 15 | 16 |
| Price ($/MWh) | 21.84 | 27.09 | 37.06 | 68.95 | 65.79 | 66.57 | 65.44 | 79.79 |
| Time (h) | 17 | 18 | 19 | 20 | 21 | 22 | 23 | 24 |
| Price ($/MWh) | 115.5 | 110.3 | 96.05 | 90.53 | 77.38 | 70.95 | 59.42 | 56.68 |

TABLE III
ASSIGNED POWER FROM THE ISO

| Time (h) | 1 | 2 | 3 | 4 | 5 | 6 | 7 | 8 |
|---|---|---|---|---|---|---|---|---|
| Power (MW) | 3.90 | 3.81 | 3.72 | 3.96 | 4.20 | 4.32 | 4.98 | 5.34 |
| Time (h) | 9 | 10 | 11 | 12 | 13 | 14 | 15 | 16 |
| Power (MW) | 5.70 | 5.94 | 6.12 | 6.24 | 6.96 | 7.92 | 8.10 | 8.28 |
| Time (h) | 17 | 18 | 19 | 20 | 21 | 22 | 23 | 24 |
| Power (MW) | 8.52 | 8.61 | 8.10 | 7.98 | 7.20 | 6.69 | 4.74 | 4.65 |

Four cases are studied as follows to show the impact of distribution grid congestion in causing a DMO cost surplus, as well as the impact of the proposed penalty factor in reducing this surplus to zero, i.e., a net-zero settlement.

**Case 1 (no congestion):** In this case, it is assumed that the line flow limits are large enough so no congestion will occur in the distribution grid. The solution of the distribution market clearing problem will result in a total customer payment of $10,850 and a total DMO payment of $10,850, this the DMO surplus will be zero. D-LMPs are calculated as the byproduct of the distribution market clearing problem which all determined to be equal to the T-LMP. This result is expected as there is no congestion in the grid to cause an increase in D-LMP values.

**Case 2 (grid congestion, no penalty):** In this case, the given capacity of the lines are considered in the market clearing problem and the proposed penalty is ignored, i.e., $\mu=0$. The results show that customer payment and the DMO payment are respectively obtained as $10,759 and $11,126. The difference between these two payments, i.e., $367, is the DMO surplus. This surplus is occurred due to the additional payment made by customers at buses 2, 11 and 12 which are facing higher D-LMP values than the rest of this system. The higher D-LMP is caused by the congestion at lines 2-3, 8-12, and 10-11. The DMO has no mechanism to redistribute this surplus to customers. If not redistributed, this surplus can potentially incentivize the DMO to operate the grid lower than the desired performance levels and seek congestion in the distribution to maximize the surplus. Thus the redistribution mechanism, as proposed in this paper, seems necessary.

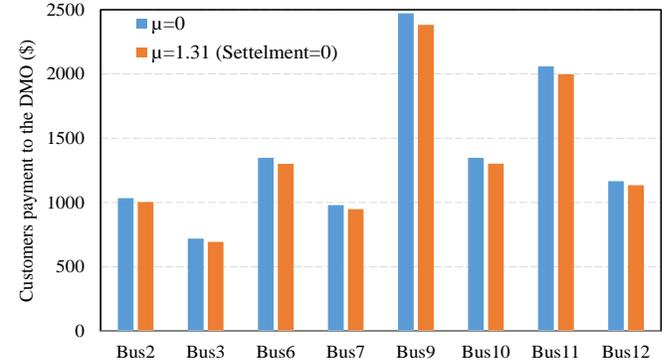

Figure 3. Customer payment to the DMO for two cases of $\mu=0$ and $C^\Delta=0$.

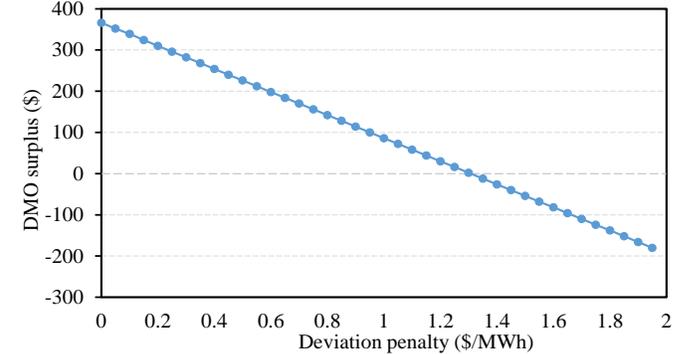

Figure 4. Net settlement as a function of deviation penalty.

**Case 3 (grid congestion, net-zero settlement):** In this case, the value of $\mu$ which leads to a net-zero settlement is calculated using the proposed model. The penalty is determined as $\mu=$1.31/MWh$. Using this penalty, the customer and the DMO payments are respectively calculated as $10,759.2 and $10,759.2, which are equal and thus ensure a net-zero settlement. Fig. 3 shows the payments of all customers, based on their location, in two cases: no penalty ($\mu=0$) and net-zero settlement ($C^\Delta=0$). The graph illustrates that the payment of all customers are decreased when considering a net-zero settlement, compared to the case with a positive surplus. This is

anticipated as the DMO surplus is redistributed among customers. The results also show that the payment reduction rate is directly proportional to the amount of the customer load.

**Case 4 (sensitivity analysis)**: The sensitivity of the distribution market settlement with respect to the penalty is investigated in this case. Fig. 4 depicts the amount of net settlement decreases by increasing the penalty. As this figure shows, the DMO cost surplus equals to zero for mismatch penalty between at $1.31/MWh, while it is positive for smaller penalty values and negative for larger penalty values.

## V. CONCLUSION

The unique structure of the distribution systems and the ongoing developments associated with the integration of DERs and responsive loads necessitate the development of proper pricing mechanisms to facilitate distribution market operation and boost distribution market participation. In this paper an analytical model was presented for calculating locational marginal prices at the distribution system as a function of distribution network congestion, individual bids of customers, T-LMP, and penalty terms. It was further investigated whether the DMO settlement can be zeroed out by having equal payments to the upstream network and from the customers. The performance of the proposed model was studied using a test distribution system.